\documentclass[aps,prm,reprint,superscriptaddress,showpacs]{revtex4-1}
\usepackage{braket}
\usepackage{graphicx}
\usepackage{dcolumn}
\usepackage{bm}
\usepackage{epsfig}
\usepackage{amsmath}
\usepackage[usenames,dvipsnames]{color}
\usepackage{booktabs}
\usepackage{hyperref}
\usepackage{color}
\usepackage[compact]{titlesec}         
\titlespacing{\section}{0.1pt}{0.1pt}{0.1pt} 
\AtBeginDocument{%
\setlength\abovedisplayskip{0pt}
\setlength\belowdisplayskip{0pt}}
\begin{document}

\title{Doping-induced magnetism in the semiconducting B20 compound RuGe} 

\author{Mojammel A. Khan}
\email{mkhan19@lsu.edu}
\affiliation{Department of Physics and Astronomy, Louisiana State University, Baton Rouge, LA 70803}

\author{D. P. Young}
\affiliation{Department of Physics and Astronomy, Louisiana State University, Baton Rouge, LA 70803}

\author{P.W. Adams}
\affiliation{Department of Physics and Astronomy, Louisiana State University, Baton Rouge, LA 70803}

\author{D. Browne}
\affiliation{Department of Physics and Astronomy, Louisiana State University, Baton Rouge, LA 70803}

\author{D. M. Gautreau}

\affiliation{Department of Physics and Astronomy, Louisiana State University, Baton Rouge, LA 70803}

\author{W. Adam Phelan}
\thanks{Current address: Department of Chemistry, The Johns Hopkins University}
\affiliation{Department of Physics and Astronomy, Louisiana State University, Baton Rouge, LA 70803}

\author{Huibo Cao}
\affiliation{Neutron Scattering Division, Oak Ridge National
Laboratory, Oak Ridge, TN 37831}

\author{J. F. DiTusa}
\affiliation{Department of Physics and Astronomy, Louisiana State University, Baton Rouge, LA 70803}

\date{\today}

\begin{abstract}
RuGe, a diamagnetic small-band gap semiconductor, and CoGe, a nonmagnetic
semimetal, are both isostructural to the Kondo insulator FeSi and
the skyrmion lattice host MnSi. Here, we have explored the magnetic and transport properties of Co-doped RuGe: Ru$_{1-x}$Co$_x$Ge.  For small values of $x$, a magnetic ground state emerges with $T_{c}\approx$ 5 $-$ 9 K, which is accompanied by a moderate decrease in electrical resistivity and a Seebeck coefficient that indicates electron-like charge carriers. The magnetization, magnetoresistance, and the specific heat capacity all resemble that of Fe$_{1-x}$Co$_x$Si for similar Co substitution levels, suggesting that Ru$_{1-x}$Co$_x$Ge hosts equally as interesting magnetic and charge carrier transport properties.
\end{abstract}
\maketitle

\section{Introduction}

Magnetic materials that lack a center of inversion symmetry often display non-collinear magnetic structures, such as the magnetic skyrmion lattice, observed in \textit{B20} systems~\citep{skyrmion.inB20,SKX.in.Cu2OSeO3}, and magnetic chiral solitons in $Cr_{1/3}NbS_{2}$~\citep{soliton.NCS.chiral}, for example. Interestingly, topological skyrmions were first predicted by Tony Skyrme in the field of high energy particle physics~\citep{skyrme.origins1,skyrme.origins2}, but they turn out to be of importance in condensed matter physics as low energy excitations
in magnetic materials that have noncentrosymmetric crystal structures
(NCS). In these magnetic compounds, the noncentrosymmetry provides the necessary condition to observe the effects of an antisymmetric exchange interaction
(also called the Dzyaloshinskii-Moriya (DM) interaction) coexisting with ferromagnetic interactions~\citep{DZYALOSHINSKY,Morya}. The competition between the symmetric exchange interaction and the anti-symmetric DM interaction results in the formation of helical magnetic spin states.  This occurs since the DM interaction is lower in energy than the isotropic exchange interaction~\citep{skyrmion.inB20}. In these materials, the skyrmion lattice structure appears just below the magnetic transition in a narrow field (\textit{H}) and temperature (\textit{T}) region~\citep{SKX.develops,phase.diagram.MnSi}.

The exotic magnetism found in transition metal silicides and
germanides having the \textit{B20} crystal structure has been of
tremendous recent interest. The most celebrated of these compounds is MnSi,
which has been characterized as a long wavelength
helimagnet~\citep{helix.MnSi}, a prototypical weak itinerant
ferromagnet~\citep{excitation.itinerant.MnSi}, a possible pressure
induced quantum critical system~\citep{NFL.itinerant.ferromagnets},
and most recently, as a host for a skyrmion lattice~\citep{muhlbauer.skyrmion}. This class of compounds also includes FeGe, Fe$_{1-x}$Co$_{x}$Si, and MnGe, all of which are helimagnets due to the significant contribution of the DM interaction. The occurrence of the skyrmion lattice phase is intimately
connected to the helimagnetism, having a characteristic wavevector that
matches the helimagnetic (HM) wavevector, $q$, despite the wide range
of $q$'s displayed (ranging from ~0.09 nm$^{-1}$ in
FeGe~\citep{lebech.magnetic.FeGe} to ~2.1 nm$^{-1}$ in
MnGe~\citep{SKX.MnGe.SANS}). The case of Fe$_{1-x}$Co$_{x}$Si is
particularly interesting, since its magnetism results from carrier
doping the small band-gap insulator, FeSi, to create a magnetic
semiconductor~\citep{manyala.FeCoSi.magnetism}. In addition, the two parent compounds, FeSi and CoSi (a diamagnetic semimetal), have no
intrinsic magnetic moments, let alone long-range order.

In this article, we report the synthesis and characterization of
the magnetic and transport properties of single crystalline
Ru$_{1-x}$Co$_{x}$Ge ($x < 0.05$) along with electronic structure calculations that largely confirm our results. RuGe is diamagnetic and a small-band gap semiconductor (band gap of $\sim$0.15 eV\citep{RuGe.band.gap}), which
crystallizes in the \textit{B20} crystal structure~\citep{transport.of.RuGe}. Interestingly, CoGe, which is isostructural to RuGe when grown
under pressure~\citep{Ditusa.paper.b20.pressure}, is also non-magnetic and a semimetal with a Dirac point just below the Fermi level~\citep{diracpoint.CoGe}. We find that
substituting Co for Ru adds magnetic moments to the diamagnetic host, which interact and result in a magnetic ground state. The dc and ac susceptibility and magnetization measurements indicate the magnetic phase transition occurs between 5 and 8.5 K for $0.02 \le x \le 0.046$. Magnetization measurements suggest a small saturation moment and a Rhodes-Wohlfarth ratio (RWR) greater than 1.
The overall behavior is akin to other \textit{B20}
compounds. Electrical resistivity measurements reveal behavior similar to that of Fe$_{1-x}$Co$_{x}$Si, where a complex resistivity consistent with a carrier-doped
semiconductor is observed, including the dominant effects of quantum interference
at low temperature.  While pure RuGe has a positive thermopower (Seebeck coefficient) at all temperatures~\citep{RuGe.band.gap}, we
find that Ru$_{1-x}$Co$_{x}$Ge has a negative thermopower.  These measurements suggest that the introduction of Co has induced a small
density of negative charge carriers, as well as magnetic moments that order below 10 K. Among the \textit{B20} compounds, Ru$_{1-x}$Co$_{x}$Ge is
only the second example, after Fe$_{1-x}$Co$_{x}$Si, in which
magnetism was found by chemical substitution between a nonmagnetic
semiconductor (RuGe) and a non-magnetic semimetal (CoGe). This is
significant, because all of the \textit{B20} materials that display
magnetic ordering host skyrmion lattices~\citep{FeGe.skx,MnGe.skx,SKX.in.Cu2OSeO3,muhlbauer.skyrmion,phase.diagram.MnSi,skyrmion.inB20}. Ru$_{1-x}$Co$_{x}$Ge may offer the opportunity to
explore these spin textures in a system, where spin-orbit coupling is enhanced, and where the properties of the spin texture are likely controlled by the level of Co substitution, as in Fe$_{1-x}$Co$_{x}$Si.
\vspace{0.2in}

\section{Experimental Description}
Single crystals of the series Ru$_{1-x}$Co$_{x}$Ge were synthesized
using a modified Bridgman technique in a radio-frequency (rf)
induction furnace. About 10 grams of elementary Ru (Alfa Aesar
99.999\%), Co (Alfa Aesar 99.99\%) and Ge (Alfa Aesar 99.999\%) were
carefully weighed, and stoichiometric mixtures for $x=0$, 0.10,
0.15, and 0.20 were placed in alumina crucibles. For each growth,
about 10\% extra Ge was added to the total weight to act as a flux. A
polycrystalline ingot was formed from this initial mixture by melting
in the rf-induction furnace. The ingot was then ground to powder and
placed in a doubly-tapered graphite crucible with a pointed
bottom. The graphite crucible was then placed inside a quartz tube and
sealed under vacuum. The tube was placed inside the rf coil suspended
by the crystal puller. Ground polycrystalline Ru$_{1-x}$Co$_{x}$Ge was
slowly melted, and then the tube was lowered through the heating zone
at ~1-1.5 mm/hr, while keeping the melt zone at constant
temperature. After 3 days of growth, the melt was found to contain
several large crystals. The melts that contained Co consisted of three
separate regions, with the lower portion containing mostly RuGe, the
middle portion consisting of Co-doped RuGe, and the top region
containing a mixture of Co-rich phases.

Attempts to grow crystals beyond the $x=0.20$ nominal Co concentration were unsuccessful, indicating a Co solubility limit at ambient pressure. This is in line with expectations, as it is known that pure CoGe only forms in the \textit{B20} crystal structure under pressure~\citep{Ditusa.paper.b20.pressure}. Because of the phase separation of the resulting materials and the apparent solubility limit, it was important to investigate the stoichiometry and homogeneity of the samples. Elemental analysis of the resulting single crystals was accomplished utilizing wavelength dispersive spectroscopy (WDS) in a JEOL JXA-8230 electron microprobe. The microprobe analysis indicated that crystals that formed in the middle portion of the melt are homogeneous in composition. Elemental analysis from the WDS measuements showed that the nominal $x=0.10$, 0.15, and 0.20 samples were Ru$_{1-x}$Co$_x$Ge with $x=0.02(1)$, $x=0.038(6)$, and $x=0.046(2)$, suggesting the solubility limit had been reached at Co substitution levels just below $x=0.05$ at ambient pressure.

These single crystals were mounted onto separate glass fiber tips using epoxy, attached to a goniometer head via the ends of brass pins, and placed on a Nonius Kappa CCD X-ray diffractometer equipped with Mo--K$\alpha$ radiation ($\lambda$ = 0.71073 \AA). The cubic Laue symmetry $m-3$ and systematic absences led to the space group selection of $P2_{1}3$ (No. 198). The generation of the initial model and subsequent structure refinement were conducted using SIR97~\citep{SIR.structure.determination} and
SHELX97~\citep{Sheldrick.crystal.xrd}, respectively. All models were corrected for extinction (SHELXL method), as well as for absorption (multi-scan method)\cite{Xrd.oscillation}. After locating all the atomic positions, the displacement parameters were refined as anisotropic, and weighting schemes were applied during the final stages of the refinement.

The crystal structure and phase purity of these crystals were also investigated by powder X-ray diffraction (PXRD) using a small portion of a powdered sample on a PANalytical Empyrean multi-stage X-ray diffractometer with Cu K$\alpha$ radiation ($\lambda$ = 1.54059 \AA). The system has a $\theta$-2$\theta$ geometry, and data were taken from $10^{\circ}$ to $90^{\circ}$ at a constant scan of $2^{\circ}$ per minute at room temperature. The data were then refined using the model determined from the single crystal XRD.

The electrical resistivity was measured using a standard four-probe ac technique, in which small diameter Pt wires were attached to the sample using a conductive epoxy (Epotek H20E).  The excitation current was between 1 and 3 mA, at a frequency of 27 Hz. Data were collected between 1.8 and 290 K and in magnetic fields up to 9 T using a Quantum Design, Physical Property Measurement System (PPMS). The specific heat was measured in the PPMS using a time-relaxation method between 2 and 50 K at 0 T. The magnetic susceptibility was also measured in the PPMS and in a Quantum Design XL-7 MPMS SQUID Magnetometer. AC susceptibility measurements were performed in the SQUID magnetometer with an excitation field of 1 Oe and a frequency of 99.99 Hz. The thermoelectric power, or Seebeck coefficient ($S$), was measured by a comparative technique in the PPMS from 2 to 350 K using a home-built sample holder with a constantan metal standard.

Single crystal neutron diffraction measurements were performed in an attempt to characterize the magnetic order of the Ru$_{1-x}$Co$_{x}$Ge crystals at Oak Ridge National Laboratory using the beamline HB3A, a four circle diffractometer, at the High Flux Isotope Reactor, HFIR.  Unfortunately, we were not able to verify the magnetic structure at the current substitution level. This indicates that the magnetic structure has a large periodicity, so that the magnetic signal overlaps with the strong nuclear Bragg scattering.  The likely large periodicity along with the fairly small saturated magnetic moment evident in the magnetization (see Sect.~\ref{mag}) combine to make the magnetic ordering difficult to observe.

\vspace{0.2in}
\section{Results and Discussions}

\subsection{Crystal Structure}
\begin{figure}[]
\centering
\includegraphics[width=0.38\textwidth]{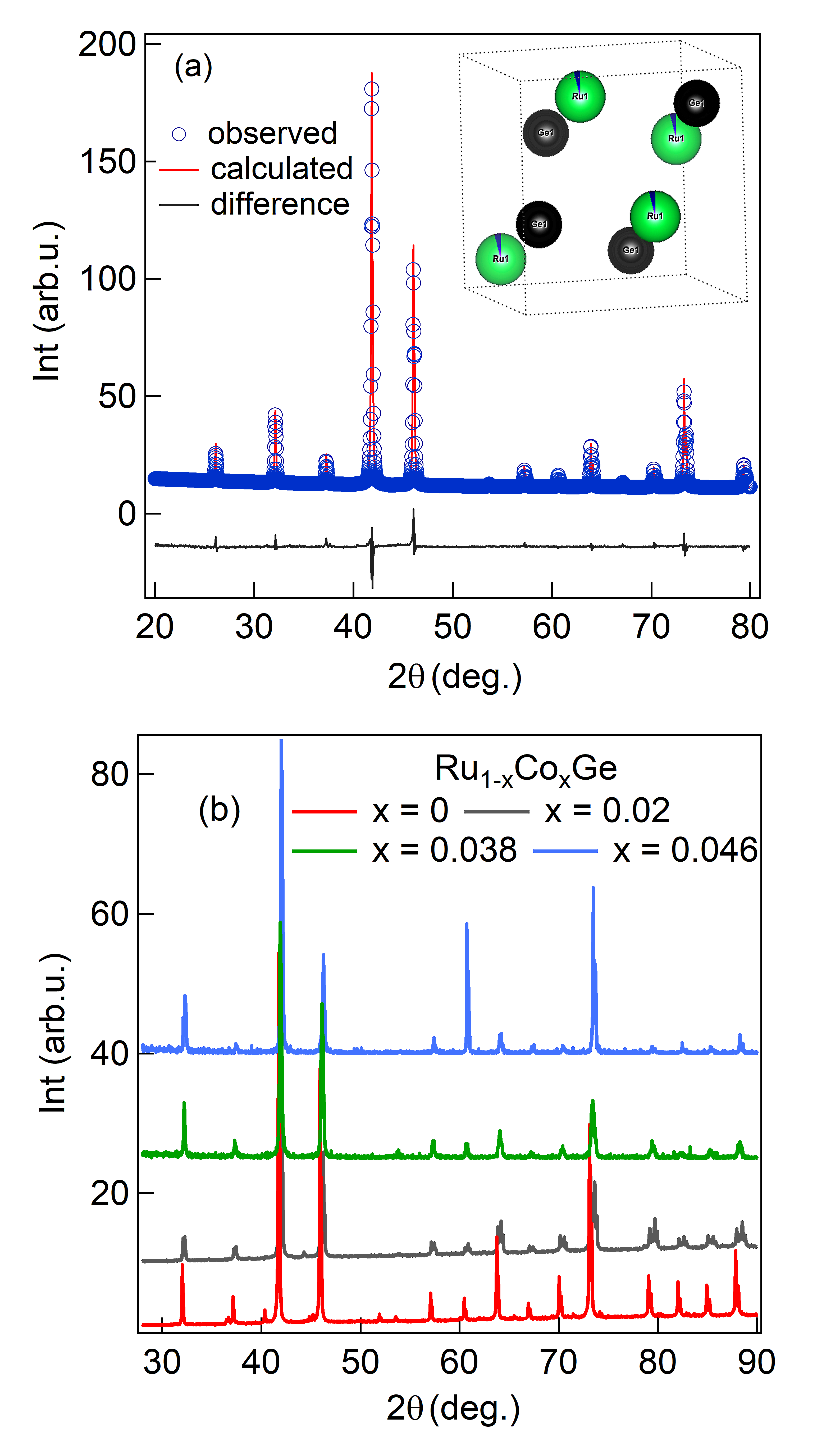}
\caption{X-ray  diffraction (XRD) investigation of Ru$_{1-x}$Co$_x$Ge. (a) PXRD of Ru$_{0.954}$Co$_{0.046}$Ge. Blue circles are the data and the red line represents the Rietveld refinement model, while the black solid line shows the difference between the data and the model. The refinement indicated the sample to have the \textit{B20} crystal structure. Inset: Schematic diagram of the crystal structure, where the green spheres represent Ru, the blue wedges represent the Co concentration on the Ru site, and the black spheres are Ge. (b) PXRD of the Ru$_{1-x}$Co$_{x}$Ge series. The red line is the PXRD data for the nominally pure RuGe crystal, which indicates the expected peak pattern for the \textit{B20} crystal structure.} \label{fig:RuCoGe powder xrd}
\end{figure}
The results of the single crystal XRD refinement are given in Table~\ref{tab:crystal structure refinement}, and a representative example of the Rietveld refinement to the powder XRD for $x=0.046$ is shown in Fig.~\ref{fig:RuCoGe powder xrd}(a). The crystal structure was confirmed to be that of nominally pure RuGe for all samples. No indications of any second phases were evident in either the single crystal or the
PXRD, as demonstrated in Fig~\ref{fig:RuCoGe powder xrd}(a \& b) for each of the samples investigated. The models refined to the single crystal XRD data were used as the starting point for the Rietveld refinements for the PXRD data. The refinements indicated a decreased lattice parameter with Co doping, as compared to that of the parent compound, RuGe, but remained larger than that of
CoGe. This is in accordance with Vegard's law and the fact that the size of the unit cell is larger for RuGe ($ a = 4.84677(3)$) than that for CoGe ($a = 4.637(3)$)~\citep{icsd}. The lattice parameters and the detailed results of the single crystal refinements for Ru$_{1-x}$Co$_x$Ge are provided in Table~\ref{tab:crystal structure refinement}.

\begin{table*}[ht]\caption{Crystallographic data for Ru$_{1-x}$Co$_{x}$Ge 
(\textit{x} =0.02, 0.038, \& 0.046)}  \centering \small
\begin{tabular}{|c|c|c|c|}
\hline
 
 & Ru$_{0.98}$Co$_{0.02}$Ge    &  Ru$_{0.962}$Co$_{0.038}$Ge &   Ru$_{0.954}$Co$_{0.046}$Ge \\
 \hline
cryst syst & cubic & cubic & cubic \\
 \hline
  space group & $P2_1$3 & $P2_1$3 & $P2_1$3 \\
 \hline
a (\AA) & 4.8330(2) & 4.8329(3) & 4. 8329(3) \\
 \hline
 \textit{V} (\AA$^{3}$) & 112.89(1) & 112.88(1) & 112.88(1) \\
 \hline
\textit{Z} & 4 & 4 & 4\\
 \hline
temp (K) & 293  & 293 & 293 \\
 \hline
& Data Collection and Refinement & & \\
 \hline
  no. of collected reflns & 400 & 321 & 322 \\
 \hline
  $\Delta \rho_{max}$ (e \AA$^{-3}$) & 1.76 & 1.78 & 1.76\\
 \hline
 $\Delta \rho_{max}$ (e \AA$^{-3}$) & - 2.27 & - 1.59 & - 1.97 \\
 \hline
 GOF & 1.15 & 1.28 & 1.14 \\
 \hline
 $R_1$(F) for $F_o^2 > 2 \sigma(F_o^2)^a$ & 0.032 & 0.029 & 0.025 \\
 \hline
 $R_w(F_o^2)^b)$ & 0.081 & 0.072 & 0.054\\
 \hline
 BASF & 0.06(3) & 0.15(2) & 0.04(3)\\
 \hline
\end{tabular} 
\label{tab:crystal structure refinement}

$^a$R$_1$(F) = $\sum\parallel F_o\mid - \mid F_c\parallel / \sum\mid F_o\mid$. $^b$R$_w$(F$_o^2$) = $\sum[w(F_o^2 - F_c^2)^2]/\sum[w(F_o^2)^2]^{\frac{1}{2}}$.
\end{table*}
\subsection{Susceptibility \& Magnetization}\label{mag}

\begin{figure}[]
\centering
\includegraphics[width=0.45\textwidth]{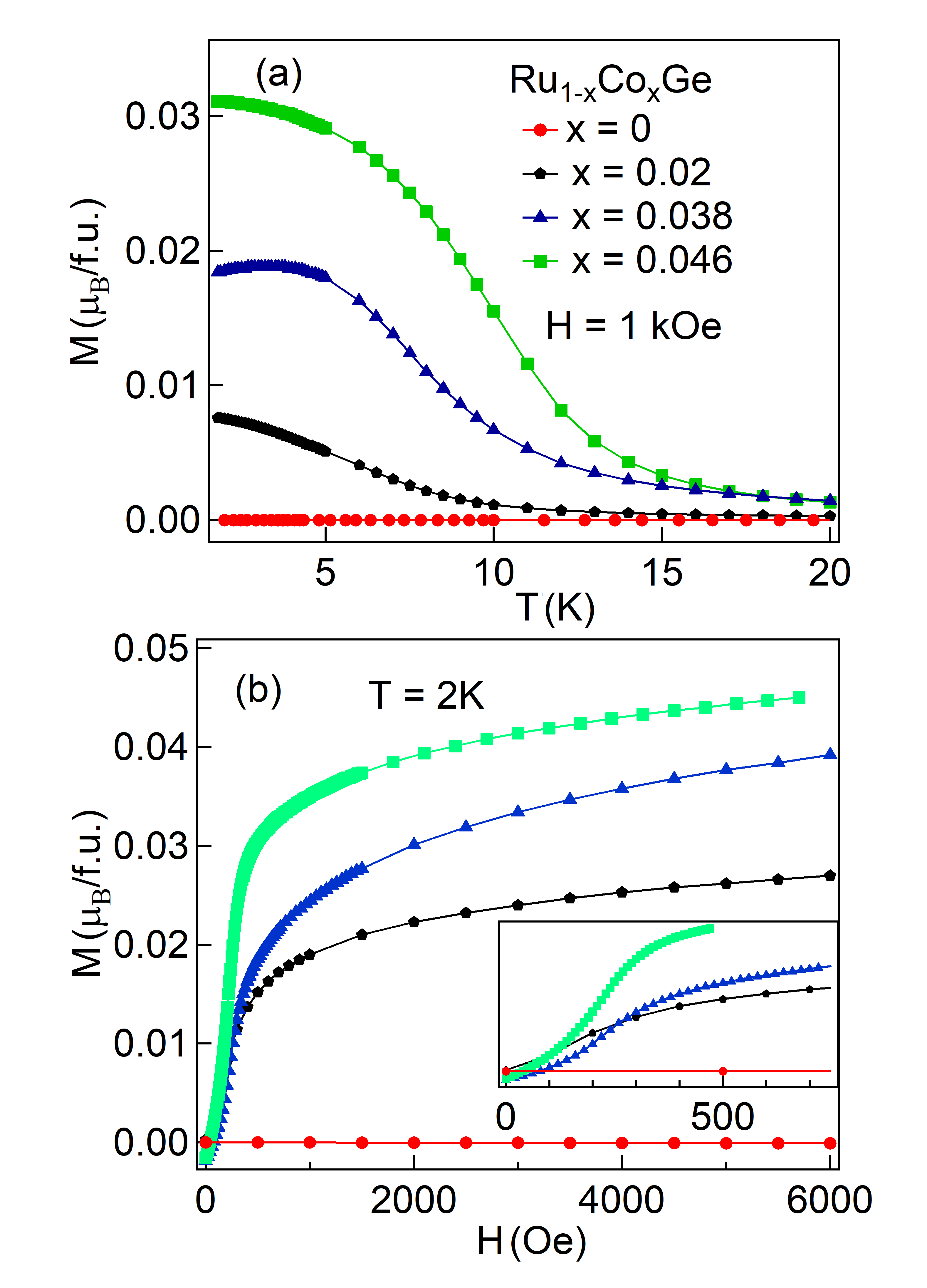}
\caption[Magnetization at 1.8 K for Co-doped RuGe]{Magnetization and
magnetic susceptibility of Ru$_{1-x}$Co$_{x}$Ge (a) Magnetization,
$M$, vs.\ temperature, $T$ in a constant field of 1 kOe. Magnetic
ordering is apparent for all crystals with Co substitution. (b) $M$
vs.\ magnetic field, $H$. Inset: Low field region of magnetization at
2 K.}
\label{fig:RuCoGe M vs H}
\end{figure}

The magnetization, \textit{M}, as a function of temperature, \textit{T}, and magnetic field, $H$, provides information on the nature of magnetic moments and their ordering. This includes the values of the fluctuating moment above, and the saturation magnetic moment below, the Curie temperature, $T_C$, as a function of Co substitution. As seen in Fig.~\ref{fig:RuCoGe M vs H}(a), the \textit{M(T)} data at $H=1$ kOe indicate a magnetic ordering, and from the minimum in $d\chi/dT$ the ordering temperature was found to be 5 K for $x=0.02$, 6.5 K for $x=0.038$, and 8.5 K for $x=0.046$. Our nominally pure RuGe crystal has a diamagnetic susceptibility with no sign of a magnetic ordering, in accordance with earlier results~\citep{transport.of.RuGe}. Thus, $T_C$ increases systematically with Co concentration, further supporting the idea that
Co substitutes for Ru at the levels indicated by the WDS and XRD data. In addition, the saturated magnetic moment increases with increasing $x$, with the largest ordered moment observed for the
Ru$_{0.954}$Co$_{0.046}$Ge crystal, as shown in Fig.~\ref{fig:RuCoGe M
vs H}(b). A large low field contribution to $M(H)$ is apparent in
Fig.~\ref{fig:RuCoGe M vs H}(b), suggesting that the magnetic order is
either ferromagnetic or helimagnetic. This result is similar to what is observed in MnGe, FeGe, MnSi, and Fe$_{1-x}$Co$_{x}$Si~\citep{helix.MnSi,Helimagnetism.FeCoSi,helimagnetism.FeCoSi.0.3to0.8,Ditusa.paper.b20.pressure}, where there is a steep increase in \textit{M(H)} at low field below $T_{C}$. Furthermore, in the magnetic \textit{B20} compounds, there is clear evidence in
\textit{M(H)} for the existence of multiple magnetic phases, including a helical ground state. As the magnetization increases, transitions from a helical to a conical phase, and then to a spin polarized phase at higher field, are observed in $M(H)$. For Ru$_{1-x}$Co$_x$Ge we observe an upturn in $M(H)$ at around 200 Oe in the
inset to Fig.~\ref{fig:RuCoGe M vs H}(b).

\begin{figure}[]
\centering \includegraphics[width=0.44\textwidth]{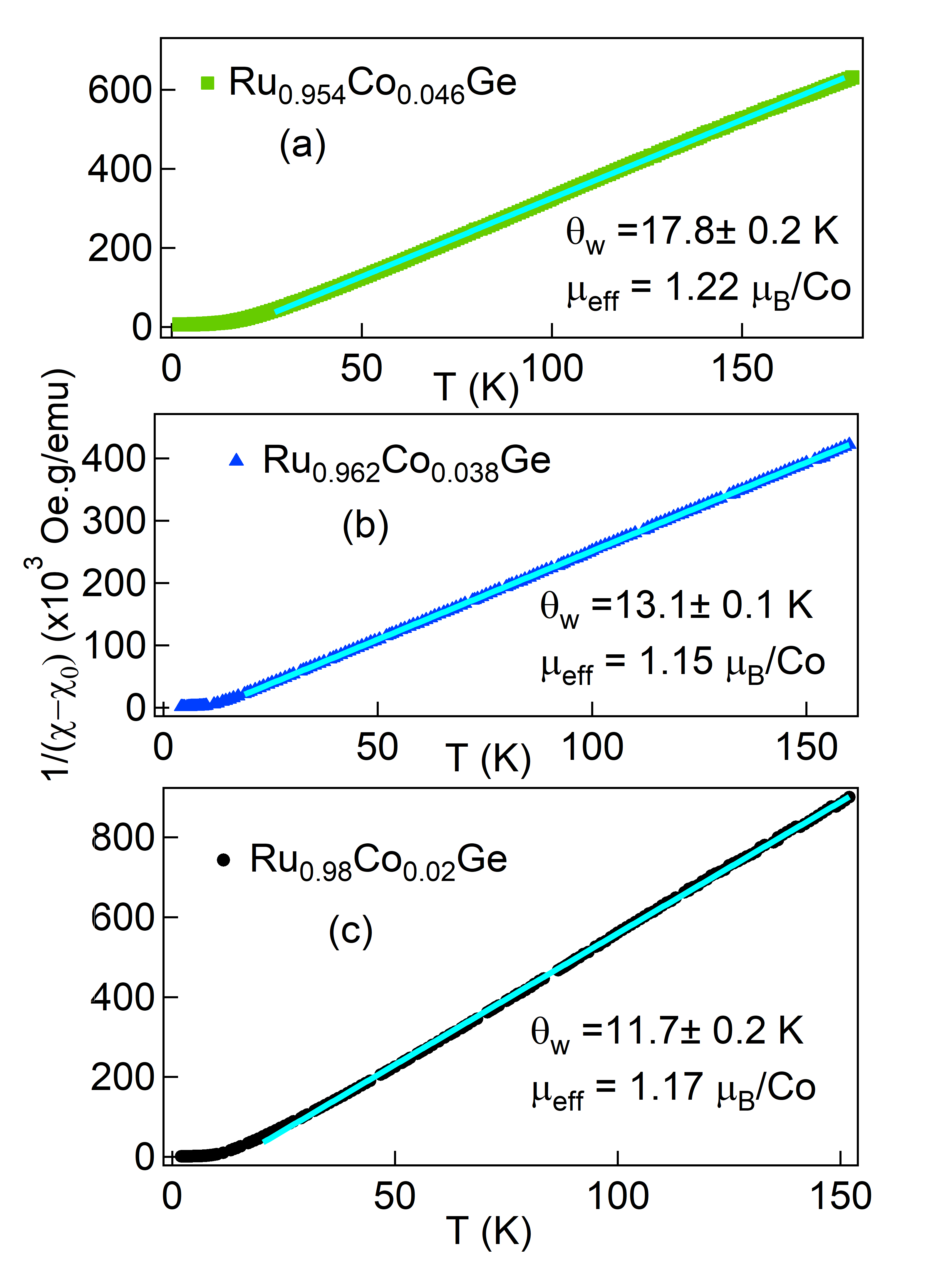}
\caption[Curie-Weiss fit to the DC susceptibility Ru$_{1-x}$Co$_{x}$Ge.] {Inverse susceptibility of Ru$_{1-x}$Co$_{x}$Ge. Inverse of the magnetic susceptibility, 1 /($\chi-\chi_0$), after subtraction of a constant paramagnetic offset ($\chi_0$). The solid line is a fit of the modified Curie-Weiss form to the data (a) for $x = 0.046$, (b) for $x = 0.03$, and (c) for $x = 0.02$.}
\label{fig:RuCoGe Curie Weiss fit}
\end{figure}

The inverse dc susceptibility, 1/($\chi - \chi_0)$ vs $T$, can be informative as fits of a Curie-Weiss form to the data indicate the size of the fluctuating magnetic moment ($\mu_{eff}$) through the Curie constant, while the Weiss temperature ($\theta_w$) provides an estimate of the magnitude and sign of the interaction between magnetic moments. Fig.~\ref{fig:RuCoGe Curie Weiss fit} presents the inverse dc susceptibility after subtraction of a small temperature independent offset, $\chi_0$, determined from fits to the data of our Ru$_{1-x}$Co$_{x}$Ge crystals, demonstrating that the modified Curie-Weiss form represents the data well for temperatures well above $T_C$. The fluctuating magnetic moment above $T_C$ increases with Co substitution, with a moment of $\sim$1.2 $\mu_{B}/Co$ for all three samples. In addition, a positive Weiss temperature, increasing with \textit{x}, indicates ferromagnetic interactions that also increase with Co substitution. The value of $\theta_w$ and $\mu_{eff}$ for each sample is indicated in Fig.~\ref{fig:RuCoGe Curie Weiss fit}. 

\begin{figure}[]
\centering
\includegraphics[width=0.44\textwidth]{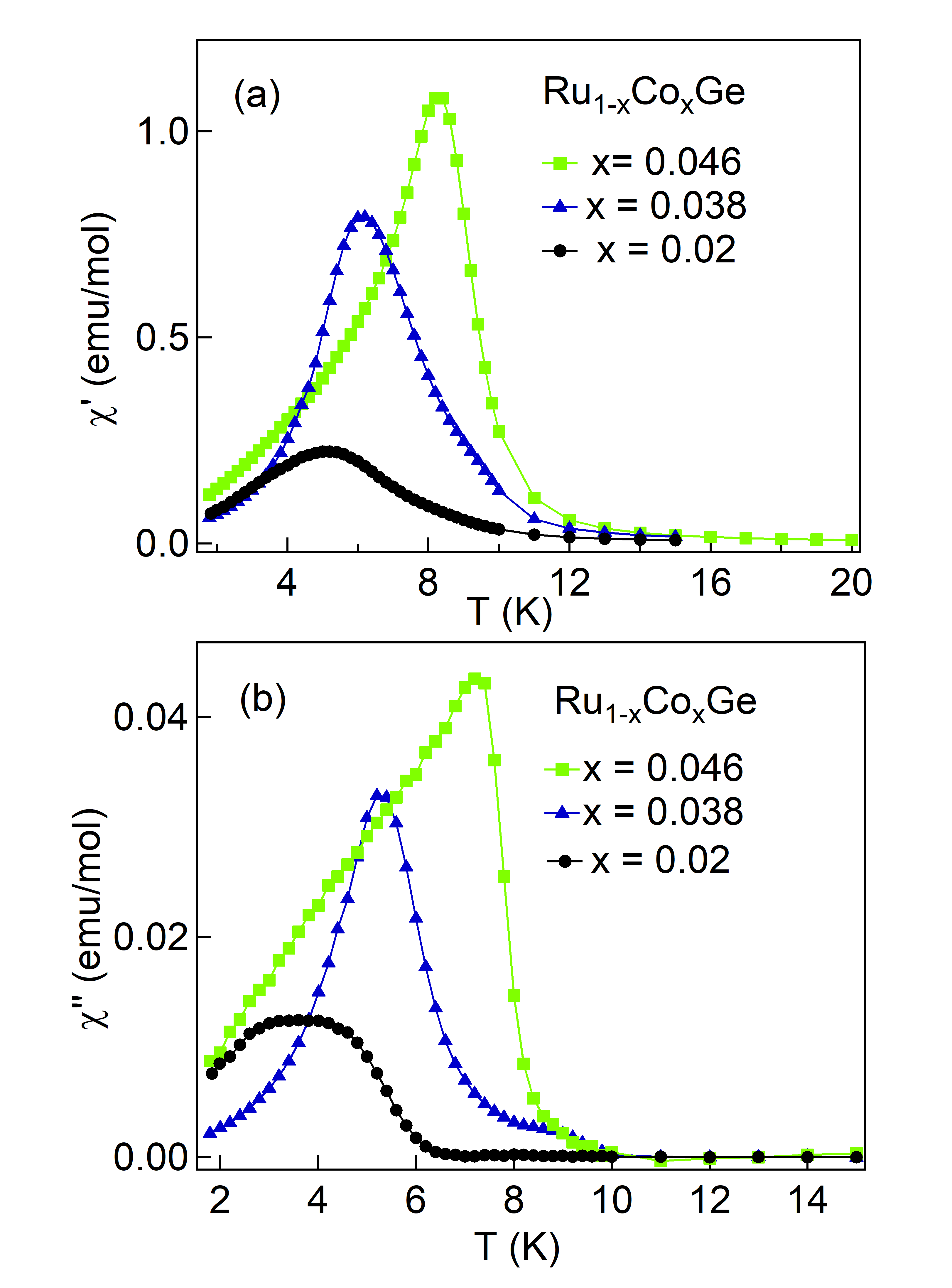}
\caption[ac susceptibility of Ru$_{1-x}$Co$_x$Ge]{ac susceptibility of
Ru$_{1-x}$Co$_x$Ge. (a) Real part of the ac susceptibility,
$\chi'$. (b) Imaginary part of the ac susceptibility, $\chi''$.}\label{fig:RuCoGe AC susceptibilty}
\end{figure}

The ac susceptibility data displayed in Fig.~\ref{fig:RuCoGe AC
susceptibilty} show clear peak structures in both the real, $\chi'$
(Fig.~\ref{fig:RuCoGe AC susceptibilty}(a)), and imaginary, $\chi''$
(Fig.~\ref{fig:RuCoGe AC susceptibilty}(b)) parts, indicating a magnetic phase
transition at these temperatures. The magnetic transition temperatures determined from the peak in $\chi'$ are within the error of those determined from the dc susceptibility, Fig.~\ref{fig:RuCoGe M vs H}. This further supports the conclusion of a magnetically ordered ground state with a small ordered moment, where both $T_C$ and the saturation magnetization increase with $x$ in Ru$_{1-x}$Co$_x$Ge.

\begin{figure}[]
\centering
\includegraphics[width=0.45\textwidth]{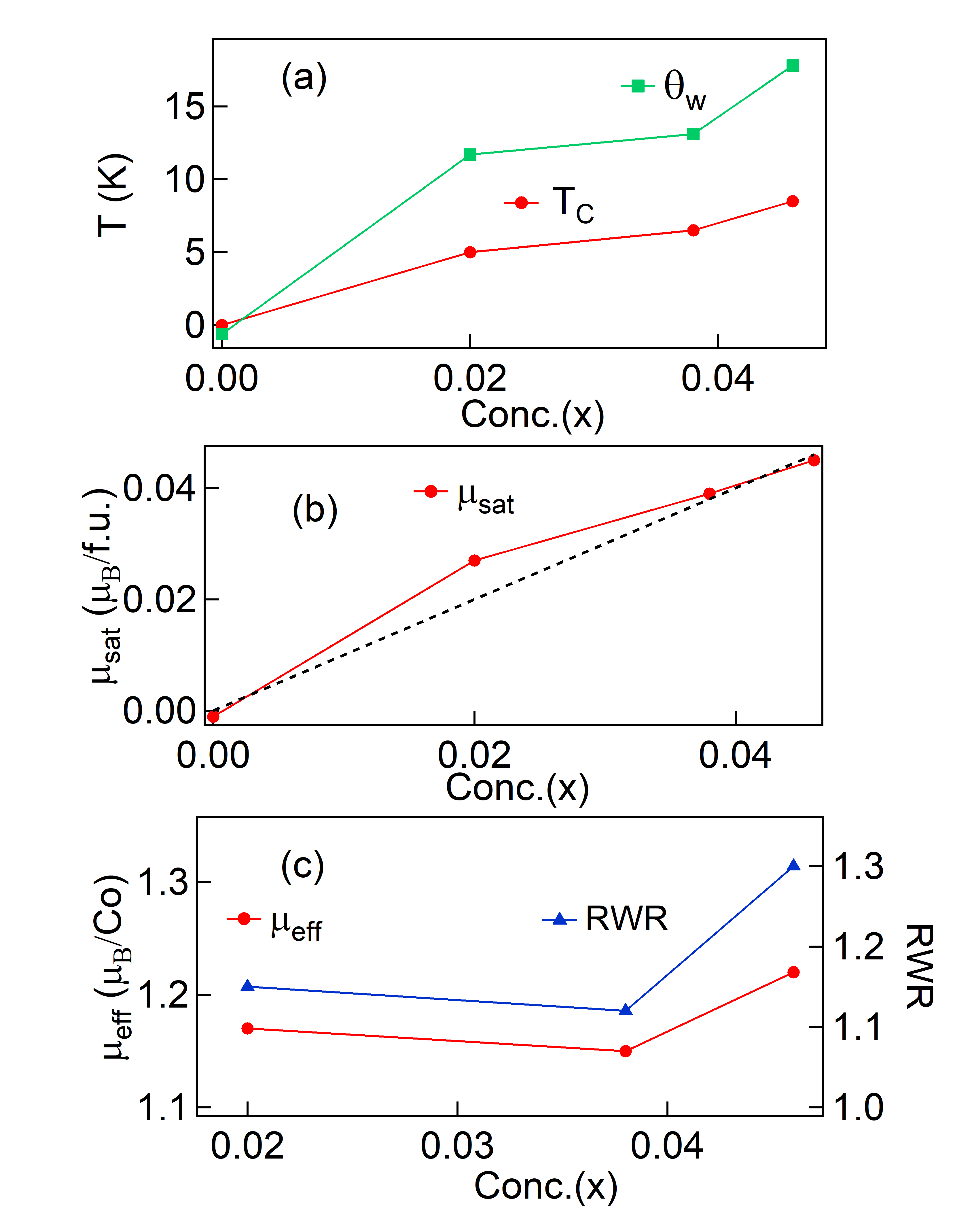}
\caption{Co concentration dependence of the magnetic parameters. (a) The Curie temperature, $T_c$, determined from the ac susceptibility and Weiss temperature, $\theta_w$, determined from a fit of the modified Curie-Weiss law are plotted vs. Co concentration, \textit{x}. (b) Saturation magnetic moment, $\mu_{sat}$ versus \textit{x}. The dashed line indicates the saturation moment of 1.00 $\mu_B/Co$. (c) The effective moment, $\mu_{eff}$, estimated from the Curie-Weiss analysis (left axis) and the corresponding Rhodes-Wohlfarth ratio (RWR) for Ru$_{1-x}$Co$_{x}$Ge (right axis).}\label{fig:RWR}
\end{figure}

Together, $M(T,H)$, $\chi'$, and $\chi''$ give the first impression of the character of the magnetically ordered state in Ru$_{1-x}$Co$_x$Ge. As previously mentioned, the apparent magnetic moment, as well as the ordering and Weiss temperatures, increase with $x$ for each of these quantities, as can be seen in Fig.~\ref{fig:RWR}. From the \textit{M(H)} data at 2 K, the saturation moment can be estimated for each of the three Co-substituted samples in Fig.~\ref{fig:RuCoGe M vs H}(b), and the values are shown in Fig.~\ref{fig:RWR}(b).

The saturated magnetic moments apparent in Fig.~\ref{fig:RuCoGe M vs H}(b) are somewhat less than the fluctuating magnetic moments found from fits to the inverse magnetic susceptibility (Figs.~\ref{fig:RuCoGe Curie Weiss fit} and ~\ref{fig:RWR}(c)). The difference between these two values indicates the degree of itinerancy in the character of the magnetism and is usually expressed in terms of the Rhodes-Wohlfarth ratio (RWR), which is the ratio of the fluctuating moment above $T_C$ to the saturation moment:  $\frac{\mu_{eff}}{\mu_{sat}}$~\citep{rhodes1963wohlfarth}. A value of
1 indicates the magnetic moments are well localized on the atomic sites, while a large RWR is associated with weak itinerant magnetism. Since the RWR is 1.3 for $x = 0.046$, 1.12 for $x = 0.038$, and 1.15 for the $x = 0.02$ sample (Fig.~\ref{fig:RWR}c), the magnetism in Ru$_{1-x}$Co$_x$Ge has a somewhat itinerant character~\cite{neutron.scattering.of.magents}. This is consistent with the itinerant nature of other transition metal \textit{B20} compounds, which have RWR values ranging from 1.32 for MnGe~\citep{MnGe.itinerant}, 2.63 for FeGe~\citep{FeGe.itinerant}, 3.5 for MnSi~\citep{rev.of.MnSi,neutron.scattering.of.magents}, and as high as 6.5
for Fe$_{1-x}$Co$_x$Si at $x=0.7$\citep{ishimoto.itinerant.Fe1-xCoxSi}. \\

\subsection{Resistivity and Seebeck Coefficient}
\begin{figure}[ht]
\centering
\includegraphics[width=0.44\textwidth]{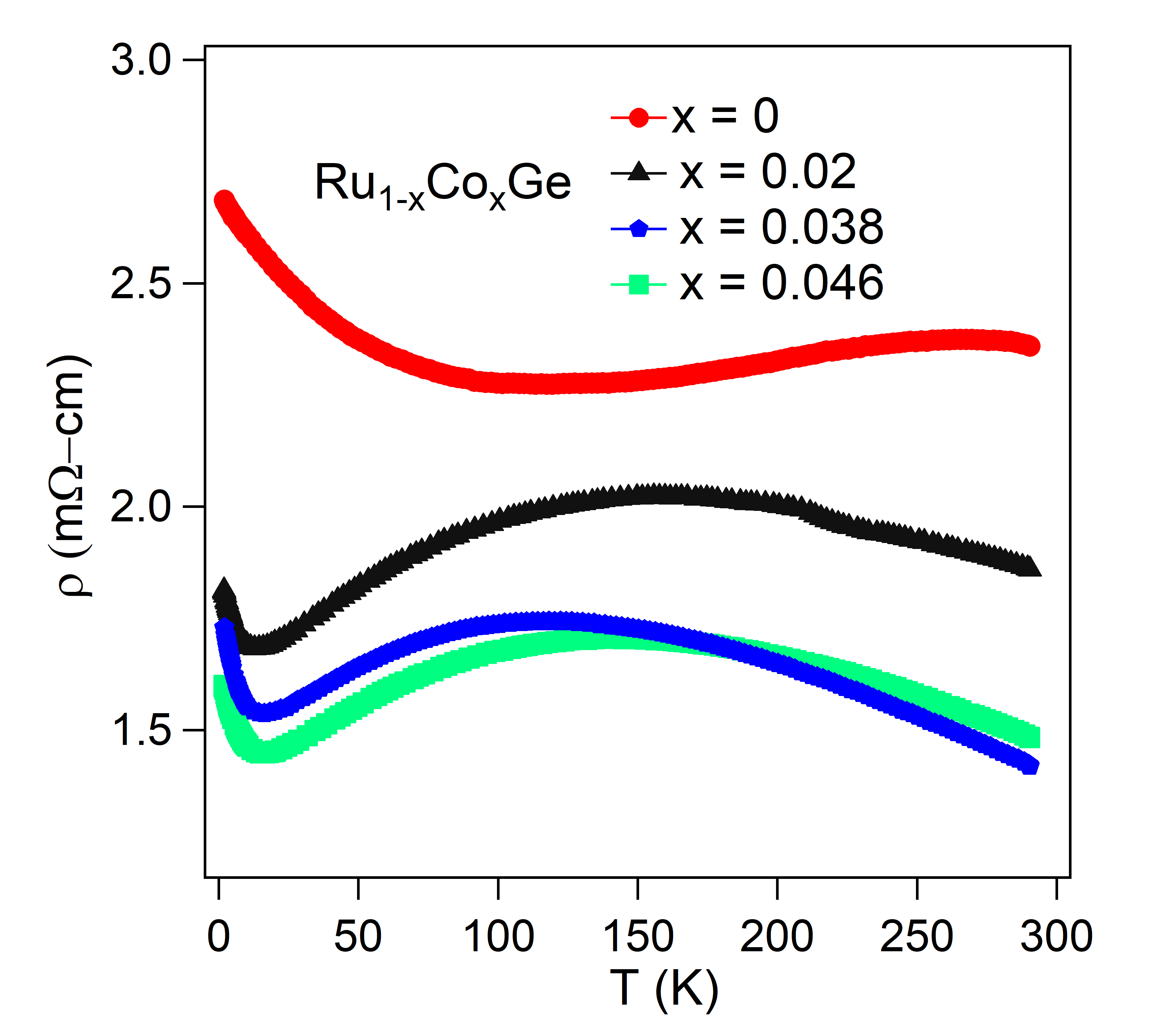}
\caption[Temperature dependent resistivity, $\rho$, of Co-doped
RuGe]{Temperature, $T$, dependence of the resistivity of Ru$_{1-x}$Co$_{x}$Ge in the temperature range of 2-290 K.}
\label{fig:RuCoGe resistivity}
\end{figure}

After establishing the magnetic ordering and the itinerant character of the magnetism in Ru$_{1-x}$Co$_x$Ge, we explored the transport properties to compare further with Fe$_{1-x}$Co$_x$Si. Previous measurements on polycrystalline arc melted samples of RuGe showed an electrical resistivity that was similar to FeSi~\citep{FeSi.Kondo}, i.e. a small-band gap insulator with an energy gap of $\sim$0.15 eV ~\citep{transport.of.RuGe}. The resistivity, $\rho$ of our nominally-pure RuGe single crystal at low $T$ is significantly smaller (by three orders of magnitude) than that of the arc melted polycrystalline samples of Ref.~\citep{transport.of.RuGe}, while the room temperature values are comparable. This indicates that the charge transport in our nominally pure RuGe crystal below 300 K is dominated by extrinsic carriers induced by defects and unintentional impurities. The Co-doped RuGe samples follow a similar trend, with a low temperature resistivity that decreases with $x$, as can be seen in Fig.~\ref{fig:RuCoGe resistivity}.

The resistivities of the Ru$_{1-x}$Co$_x$Ge samples shown in
Fig.~\ref{fig:RuCoGe resistivity} are similar in form and magnitude to
those of Fe$_{1-x}$Co$_{x}$Si~\citep{proeprties.of.Fe1-xCoxSi} for similar levels of Co substitution, but distinct from that of non-magnetic FeSi$_{1-z}$Al$_z$~\citep{FeSi.Al.doped}. In the region around 150 $\sim$ 290 K, $\rho$ decreases with decreasing temperature, which is likely due to either thermal activation above the small band gap of RuGe or carrier hopping. In the intermediate temperature regime below 100 K, the resistivity behavior is consistent with that of a poor metal with $\frac{d\rho}{dT}>0$. The upturn in the resistivity at low temperature, below about 12 K in the Co-substituted crystals, and below 100 K in our nominally pure RuGe crystal, is likely due to electron-electron (e-e) interactions, which are common among carrier-doped semiconductors~\cite{quantum.interference.transport1} in proximity to an insulator-to-metal transition. In this disordered transport regime, charge carriers have a short mean-free path, which creates highly diffusive motion. However, at low temperatures, the time associated with phase breaking can be much longer than the mean scattering time. Here, Coulomb interactions are effectively amplified by the probability of two carriers interacting more than once within a phase-breaking scattering time, which leads to coherent quantum interference of the scattering amplitudes~\citep{e.e.interaction1,e.e.interaction2,e.e.interaction3}. The result is an enhanced Coulomb coupling and a square-root singularity in the electronic density of states at the Fermi level~\citep{manyala.FeCoSi.magnetism,e.e.interaction1,e.e.interaction2}. This singularity is manifested in the transport behavior, such as the temperature-dependent resistivity and magnetoresistance, which persists to low (even zero) temperature.

The resistivity as a function of temperature, $\rho(T)$, is strikingly similar to that of
Fe$_{0.95}$Co$_{0.05}$Si, where e-e interactions in the presence of the finite sample magnetization was found to be the dominant contributor to the low
$T$ $\rho(T,B)$ ($B=H + \alpha M$)~\citep{Metal.insulator.transition.FeSi,proeprties.of.Fe1-xCoxSi}.
Just as in Fe$_{1-x}$Co$_x$Si, the upturn in the resistivity occurs at a slightly higher temperature than $T_C$, where locally $M(H=0)$ becomes non-zero.

\begin{figure}[ht]
\centering
\includegraphics[width=0.44\textwidth]{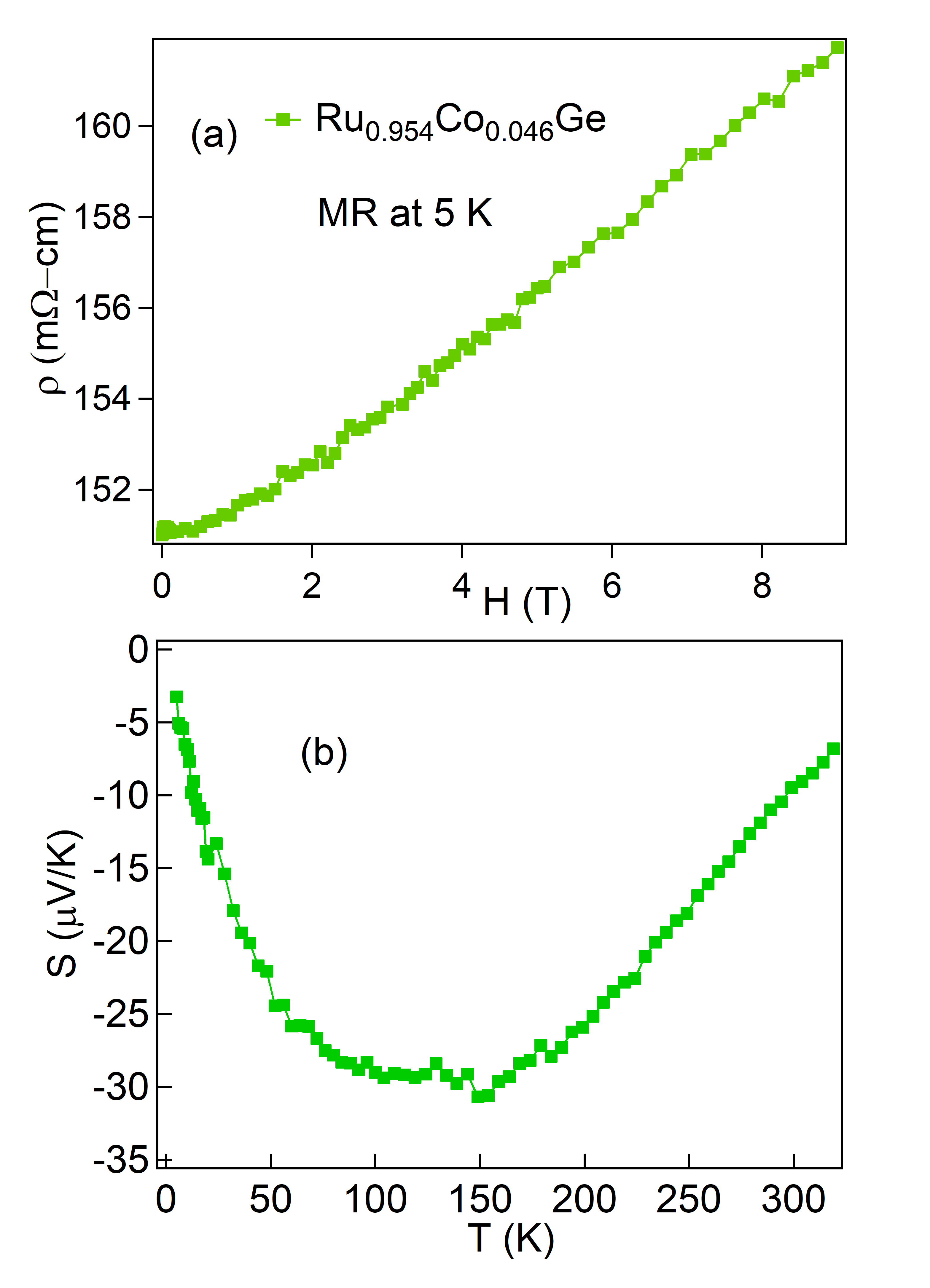}
\caption[Magnetoresistance at 5 K of the 20\%
sample]{Magnetoresistance and thermoelectric power. (a)
Magnetoresistance at 5 K of Ru$_{0.954}$Co$_{0.046}$Ge. (b)
Seeback coefficient of Ru$_{0.954}$Co$_{0.046}$Ge.}
\label{fig:RuCoGe MR}
\end{figure}

The magnetoresistance (MR) for the $x=0.046$ sample shown in
Fig.~\ref{fig:RuCoGe MR}(a) is representative of that for all the samples
and is positive across the entire field range, with a quadratic
field dependence evident at low field. Again, this behavior is similar to
that of Fe$_{1-x}$Co$_{x}$Si, where the e-e interactions dominate the
MR leading to a square-root like dependence at higher field~\citep{proeprties.of.Fe1-xCoxSi,manyala.FeCoSi.magnetism}. This is interesting because e-e interactions are known to dominate the charge transport of common semiconductors, such as Si and Ge. Finding similar behavior in magnetic semiconductors, such as the Ru$_{1-x}$Co$_x$Ge series reported here, emphasizes that the transport in different types of semiconductors can be understood from similar principles.

The thermopower, or Seebeck coefficient, is an important property of semiconductors and is of interest from a practical standpoint for potential applications in electronic refrigeration and power generation~\citep{thermoelectric.power2,thermoelectric.power}. From a measure of the thermopower, the sign of the majority carriers can be determined. The Seebeck coefficient, $S$, for pure RuGe was previously found to resemble that of an intrinsic semiconductor with $p$-type conduction, i.e. the majority carriers are holes with a positive thermopower~\citep{transport.of.RuGe}. In contrast, the
Ru$_{0.954}$Co$_{0.046}$Ge, as displayed in Fig.~\ref{fig:RuCoGe
MR}(b), exhibits a negative Seebeck coefficient, and is therefore an $n$-type semiconductor. This is to be expected from the simple picture of a semiconductor being doped with an atom that has a greater number of valance electrons (Co having one more $d$-electron than Ru). With decreasing temperature, the Seebeck coefficient displays a broad minimum and then increases toward zero at zero temperature. The room temperature value of $S$ for Ru$_{0.954}$Co$_{0.046}$Ge (-10 $\mu$ V/K) is smaller in magnitude and opposite in sign, as compared to that of the previously measured nominally pure RuGe (+27 $\mu$ V/K)~\citep{transport.of.RuGe}. Similar behavior was also observed in Fe$_{1-x}$Co$_{x}$Si~\citep{proeprties.of.Fe1-xCoxSi}.

\subsection{Heat Capacity}

\begin{figure}[]
\centering
\includegraphics[width=0.44\textwidth]{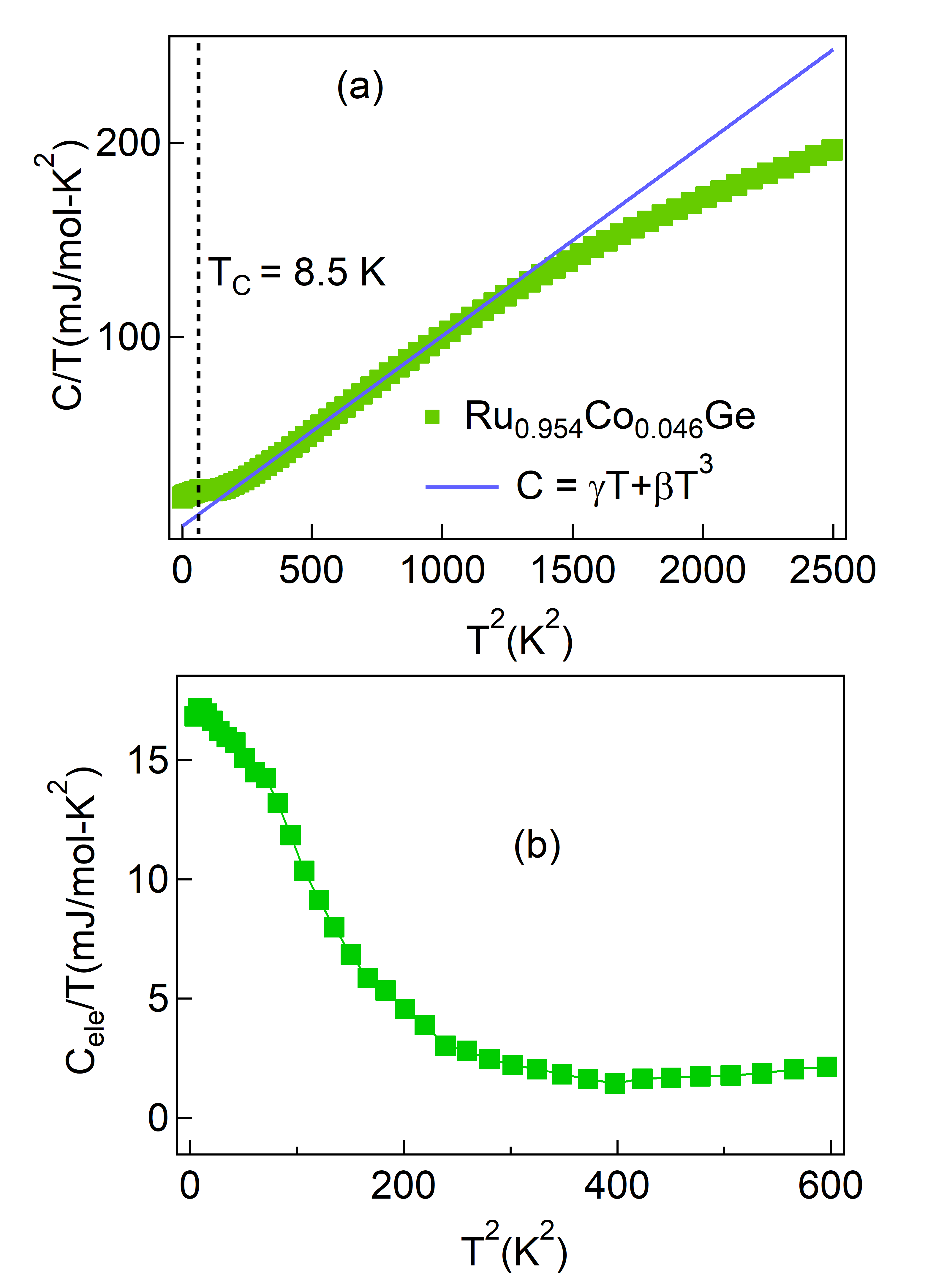}
\caption[Heat capacity of doped RuGe as a function of temperature]{Specific heat capacity of Ru$_{0.954}$Co$_{0.046}$Ge. (a) The total specific heat capacity at constant pressure, $C$ of Ru$_{0.954}$Co$_{0.046}$Ge divided by temperature, $T$ vs.\ $T$. The solid line is a fit to the data using the standard relationship for metals (see text). The dashed line indicates the Curie temperature (8.5 K), where a maximum in $C/T$ is observed. (b) The electronic heat capacity calculated after subtraction of the phonon contribution determined from the fit to the data in (a).}
\label{fig: RuCoGe heat capacity}
\end{figure}

The specific heat capacity, $C$, of Ru$_{0.954}$Co$_{0.046}$Ge is shown in Fig.~\ref{fig: RuCoGe heat capacity}. A maximum in $C/T$ coincides with the magnetic ordering temperature identified earlier in magnetization and susceptibility data (Figs.~\ref{fig:RuCoGe M vs H} \& \ref{fig:RuCoGe AC susceptibilty}), verifying that a bulk magnetic transition occurs at this temperature. The specific heat above 14 K was fit with the standard form for a metal, $C/T =\gamma + \beta T^2$, with the best fit displayed in Fig.~\ref{fig: RuCoGe heat capacity}(a). The fit yields estimates of the Sommerfeld coefficient, $\gamma$ = 2.2 $\pm$
0.5 mJ mol$^{-1}$ K$^{-2}$ and the Debye coefficient $\beta$ = 0.0984
$\pm$ 0.0004 mJ mol$^{-1}$ K$^{-4}$. The Debye temperature can be estimated from the relation $\theta_{D} = (\frac{5 \beta}{N12\pi^{4}
R})^{-1/3}$, yielding $\theta_{D}$ = 430 K. The value for $\gamma$ that we find is large considering the likely small density of carriers induced by the Co substitution. However, it is comparable to, but only about half as large as that found for similar Co doping levels in FeSi\citep{increasing.gamma.with.dopingFesi,Increasing.gamma2}. In addition, the entropy associated with the magnetic ordering can be estimated by the determination of the area between the data of Fig.~\ref{fig: RuCoGe heat capacity}b and a constant offset equal to $\gamma$. This procedure results in a value of $S_{mag}$=0.22$\pm$0.04 J mol$^{-1}$ K$^{-1}$ which can be compared to estimates based on simple assumptions. If we assume that each Co substituted into RuGe contributes a spin-1/2 magnetic moment that orders below $T_C$, we can estimate the magnetic entropy expected using the relation~\citep{gopal2012specific}, $S_{mag} = xRln(2)$, where \textit{R} is the molar constant. For $x=0.046$  the expectation is $S_{mag}$=0.26 J mol$^{-1}$ K$^{-1}$ which is within error of our $S_{mag}$ determined from the specific heat data.

\vspace{0.1in}
\subsection{Electronic Structure}
We calculated the electronic structure of RuGe via density functional theory (DFT) using the Perdew-Burke-Ernzerhof GGA functional~\citep{dos1}, employing the linear augmented plane wave plus local orbital (LAPW+LO) basis implemented in the WIEN2K software package~\citep{dos2}. Spin orbit interactions were included in the functional. The calculations for RuGe employed a lattice constant of 4.846 \AA, with atomic positions (\textit{u,u,u}), where \textit{u}=0.1327 for Ru and \textit{u}=0.8384 for Ge. These values were the equilibrium positions for the atoms calculated with DFT. The muffin tin radii for Ru was 2.38 Bohr, and that of Ge was 2.27 Bohr. The Brillouin zone integrations used a 34$\times$34$\times$34 grid of \textit{k}-points. The plane wave cutoff was varied, with convergence appearing at R*K=8.

\begin{figure}[]
\centering
\includegraphics[width=0.42\textwidth]{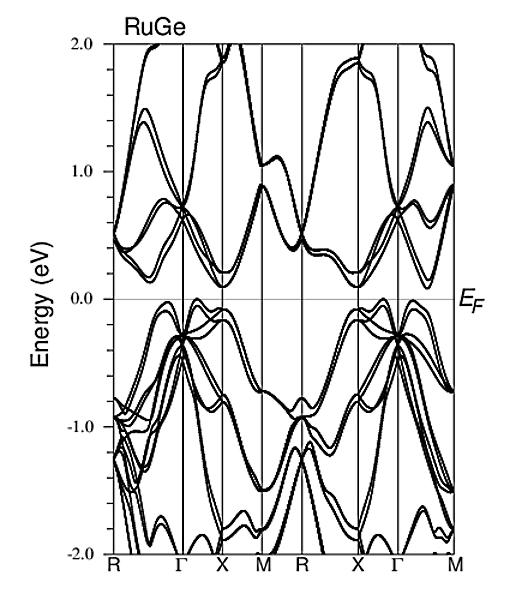}
\caption{Electronic Band structure of RuGe. Section of the calculated electronic band structure of RuGe along high symmetry directions within the energy $\pm$2 eV around the $E_F$. Spin orbit coupling was included in the calculation.}
\label{fig: RuGe Band}
\end{figure}

\begin{figure}[]
\centering
\includegraphics[width=0.48\textwidth]{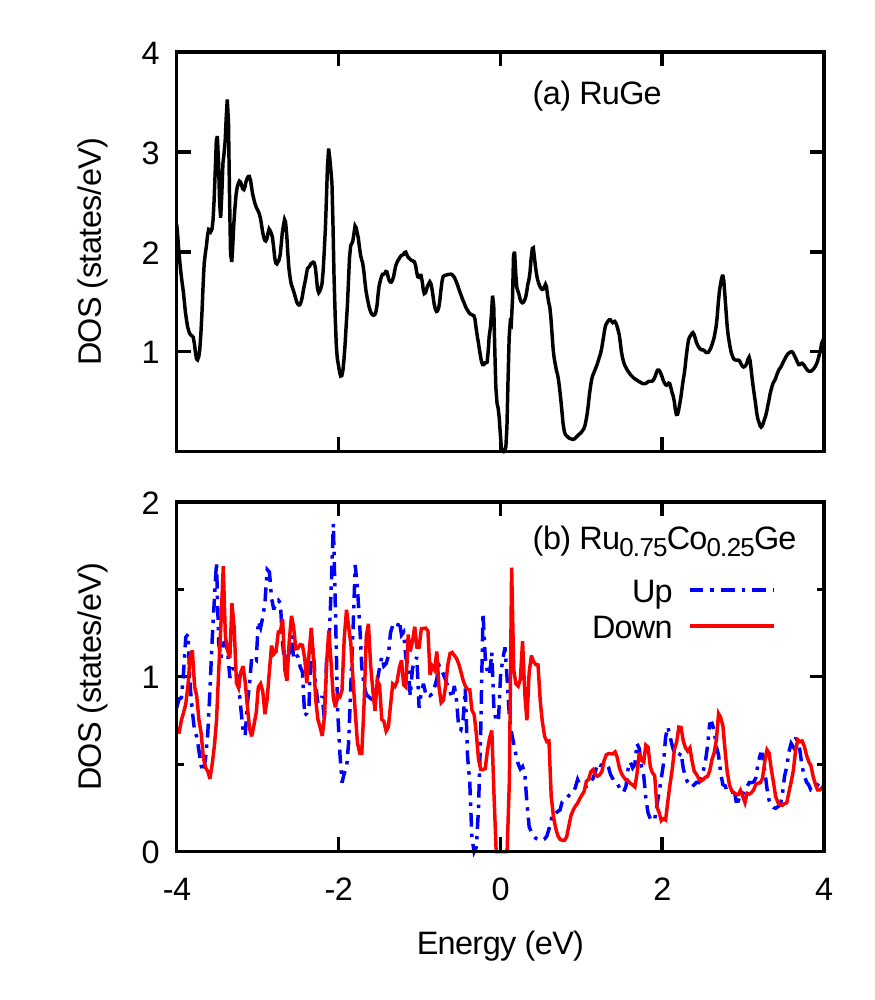}
\caption{Density of States (DOS) of pure and Co substituted RuGe. (a) DOS of RuGe. (b) Ru$_{0.75}$Co$_{0.25}$Ge, where the majority spin sub-band DOS is indicated by the dashed-dotted line (blue) and the minority spin sub-band DOS represented by the solid line (red).}
\label{fig:RuCoGe DOS}
\end{figure}

The calculated electronic band structure is shown in Fig.~\ref{fig: RuGe Band} with the total DOS displayed in Fig.~\ref{fig:RuCoGe DOS}. RuGe appears to be a narrow ($\sim$0.18 eV) indirect gap semiconductor, with the bands in proximity to the Fermi level composed almost entirely of Ru \textit{d} orbitals. As has been predicted for other B20 structured materials such as CoGe~\citep{diracpoint.CoGe} and CoSi~\citep{diracpoint.cosi}, RuGe displays two Dirac points at $\Gamma$, one about 0.25 eV below the valence band maximum, and the other about 0.6 eV above the conduction band minimum. No attempt was made to improve the band gap estimate by employing a Becke-Johnson correction~\citep{dos3} to the conduction band energies. These results are comparable to an earlier calculation of the electronic structure of RuGe~\citep{RuGe.structure}.

Attempts to use a 2$\times$2$\times$2 supercell with one of the 32 Ru atoms replaced with a Co atom to simulate Ru$_{1-x}$Co$_x$Ge with \textit{x}$\sim$0.03 did not conclusively determine whether the ground state was magnetic. For these calculations, we used a smaller lattice constant of 4.833 \AA, and a Co muffin tin radius of 2.30 Bohr.  The smaller unit cell allowed us to use a smaller 19$\times$19$\times$19 grid in the Brillouin zone. The energy difference between the magnetic and non-magnetic states was close to the resolution of the calculation, with the magnetic state slightly lower in energy. However, the resolution of the calculation is not sufficient to determine that the magnetic state is the ground state of this system. 

As an alternative, we investigated the case of \textit{x} = 0.25 with a single unit cell and one Ru replaced with Co. The resulting density of states is displayed in Fig.~\ref{fig:RuCoGe DOS}. To account for the smaller size of the Co ion, we adopted a lattice constant of 4.795 \AA.  This material is half metallic with the band gap persisting in the minority spin-sub band. The total magnetic moment is 1 $\mu_B$ per formula unit, with 0.67 $\mu_B$ on the Co and 0.14 $\mu_B$ on each of the Ru with a small negative contribution on the Ge sites and spread out over the unit cell emphasizing an itinerant nature of the magnetism. The energy difference between the paramagnetic and ferromagnetic state was 40 meV. The result of this calculation is similar to the case of Fe$_{1-x}$Co$_x$Si, where a half metallic state was found for $x<0.25$~\citep{guevara2002electronic}, and is consistent with the experimental results, where low levels of Co substitution for Ru results in a magnetic ground state with roughly 1 $\mu_B$ per formula unit. 

The calculation that we have performed for \textit{x} =0.25 also suggests that the rigid band approximation is applicable for doped RuGe with little change in the shape of the DOS from the pure RuGe. The extra carriers introduced by Co substitution fill the majority band, leading to half metallic behavior and an effective moment of 1 $\mu_B/Co$ and the total magnetic entropy amounts to $0.25Rln(2)$ per formula unit for this doping level. Since the DOS per spin is 0.97$\pm$0.05 states/eV near the Fermi level, we expect a value for $\gamma$ of 2.3$\pm$0.1 mJ mol$^{-1}$ K$^{-2}$. These predictions are similar to what has been observed experimentally at \textit{x} = 0.046
from the heat capacity data. The similarity between our data for x =0.046 and calculation for \textit{x} = 0.25 suggests that at lower doping, nearly all of the carriers contributed by the Co substitution reside in the conduction band of the majority spin. The majority spin sub-band is shifted by exchange to create a half metallic material. Because of the relatively constant DOS of 0.9$\pm$ 0.1 states/eV/spin for the conduction band near the Fermi level, we would predict the electronic contribution to the specific heat, $\gamma$, would remain in the 2.2 to 2.4 mJ mol$^{-1}$ K$^{-2}$ range at low temperature with each Co contributing one charge carrier and an ordered magnetic moment of 1 $\mu_B$ independent of \textit{x}.

\vspace{0.1in}
\section{Conclusions}
In an effort to discover new materials that could possibly host emergent magnetic phenomena, such as a magnetic skyrmion lattice, we have synthesized and explored the magnetic and transport properties of single crystals of Ru$_{1-x}$Co$_{x}$Ge. We established that Co substitution takes place, but at a level significantly below that of the nominal values from our starting materials. In addition, a solubility limit was established at about $x=0.05$ for ambient pressure synthesis, which is not unexpected, as the end member CoGe only forms in the \textit{B20} structure under high pressure. Magnetic ordering was induced by Co substitution in all of our samples for ($0.02 \le x\le 0.046$). These results are consistent with electronic structure calculations performed with one fourth of the Ru replaced with Co within the unit cell of RuGe. Thus, we have discovered that mixing the nonmagnetic small-gap semiconductor RuGe with the nonmagnetic semimetal CoGe has created a ferromagnetic semiconductor, despite the lack of magnetic moments in either RuGe or CoGe.
Exploring the charge carrier transport and magnetic properties at lower $x$ in Ru$_{1-x}$Co$_x$Ge will be an important extension of this work, since non-Fermi liquid behavior was discovered in the related series Fe$_{1-y}$Mn$_y$Si near the insulator-to-metal transition. This behavior was shown to be due to the underscreening of magnetic moments via the Kondo effect~\citep{quantum.interference.transport2}. This is significant because of the importance of, and variety of effects due to, the interactions between the magnetic and charge degrees of freedom in magnetic semiconductors having the $B20$ crystal structure.

This work demonstrates that the presence of Fe and Mn is not necessary for
nucleating a magnetic state in an FeSi-like cubic system.  More importantly, the magnetism in Ru$_{1-x}$Co$_x$Ge is likely helimagnetic, as it is in all of the transition metal $B20$ compounds that magnetically order, making this material a good candidate for skyrmion lattice formation.  We would expect the helimagnetic wave vector, \textit{q}, to be determined by a larger spin-orbit coupling, simply because of the larger atomic masses, compared to that of the silicides, FeGe, or MnGe. Naively, this would argue for a larger \textit{q} that varies with \textit{x} in a similar fashion to the case of Fe$_{1-x}$Co$_{x}$Si~\citep{Helimagnetism.FeCoSi}. 
If indeed this is the case, then Ru$_{1-x}$Co$_{x}$Ge will be a
germanide counterpart to FeSi, and the presence of another material demonstrating skyrmion lattice formation will be important for establishing the necessary energy scales and the physics behind the DM interaction. Recently, it has been suggested that features of the electronic structure, rather than simple geometry of the crystal structure, determines the strength of the DM interaction~\citep{controlDMI}. Hence, systems where the Fermi level and density of electronic states can be controlled over a wide range via chemical substitution are important for exploring the validity of this hypothesis. Thus, this work will serve as impetuous for synthesis of higher concentrations of Co-doped RuGe compounds,
in which the realization of skyrmion phase can be investigated.

\vspace{0.2in}

\acknowledgments

 JFD and DPY acknowledge the support of the U.S. Department of Energy
(DOE) under EPSCoR Grant No. DE-SC0012432, with additional support from the Louisiana Board of Regents for crystal structure determination and characterization. D.P.Y. acknowledges support from the NSF under Grant
No. DMR-1306392 for synthesis and materials. P.W.A. acknowledges support by the DOE under Grant No. DE-FG02-07ER46420 for thermodynamic measurements. We also acknowledge Dr. Clayton Loehn at the Shared Instrumentation Facility (SIF), LSU for chemical analysis.  The work at ORNL's High Flux Isotope Reactor was sponsored by the Scientific User Facilities Division, Office of Science, Basic Energy Sciences, U.S. DOE.

\bibliography{mybib}

\end{document}